\documentstyle[prb,aps,epsfig]{revtex}

\tightenlines
\begin{document}
\title{Exact-exchange density-functional calculations for
noble-gas solids}
\author{R.~J.~Magyar$^1$, A.~Fleszar$^2$, and E.~K.~U.~Gross$^{1,2}$}
\address{
$^1$Theoretische Physik, Freie Universit\"at Berlin, Arnimallee 14,D-14195
Berlin, Germany\\[1ex]
$^2$Theoretische Physik I, Universit\"at W\"urzburg, Am Hubland, D-97074
W\"urzburg,
Germany}

\maketitle
\begin{abstract}
The electronic structure of noble-gas solids is calculated
within density functional theory's exact-exchange method (EXX) and
compared with the results from the local-density approximation (LDA).
It is shown that the EXX method does not reproduce the fundamental
energy gaps as well as has been reported for semiconductors. However,
the EXX-Kohn-Sham energy gaps for these materials reproduce about 80 \%
of the experimental optical gaps.
The structural properties of noble-gas solids are described by the EXX 
method as poorly as by the LDA one. This is due to missing Van der Waals 
interactions in both, LDA and EXX functionals.

\end{abstract}
\vspace*{2cm}
PACS: 71. 71.10.-w  71.20.Nr
\newpage
\section{Introduction}

Density Functional Theory (DFT) is nowadays by far the most popular \emph{ab
initio}
method to calculate ground state properties of atoms, molecules and
solids.\cite{HK64,KS65,DG90}
Its generalization to time-dependent phenomena\cite{RG84} extended the 
scope 
of applied DFT to electronic excitations. DFT's successes rely on the
construction
of accurate and practical approximations to the
exchange-correlation part of the energy functional. The local-density
approximation (LDA) proposed at the very beginning of DFT\cite{HK64,KS65} was 
and remains a simple tool to use, whose accuracy, however,
is surprisingly good given the simplicity of the approximation.
Over the years, LDA has been applied to practically all systems
of interest.
In most cases, LDA describes ground state properties, such as
equilibrium structures, adiabatic phonons, to mention only a few,
surprisingly well, 
even for materials with strongly inhomogeneous electron densities.
On the other hand, attempts to describe
energy gaps, E$_g$, or electronic excitations brought rather limited
success.

The case of E$_g$ is a special one because E$_g$ is defined
as the difference between the ionization potential, $I$, and
electron affinity, $A$. Hence, E$_g$ can be written in terms of total 
ground-state
energies of systems with different numbers of electrons, so it is in
principle a ground state property:
\begin{equation}
{\rm E}_g = I - A = E[N+1] + E[N-1] - 2 E[N].
\end{equation}
Here $E[N+1]$, $E[N]$, and $E[N-1]$ are the total ground-state energies
of the systems with $N+1$, $N$ and $N-1$ electrons respectively. It has
been shown\cite{PL83,SS83} that E$_g$ can be re-expressed
in the following form:
\begin{equation}
{\rm E}_g = {\rm E}_g^{KS} +\Delta_{xc},
\end{equation}
where
\begin{equation}
{\rm E}_g^{KS} = \epsilon_{N+1}(N) - \epsilon_N(N),
\end{equation}
is the Kohn-Sham (KS) gap and $\Delta_{xc}$ is the discontinuity in the
exchange-correlation potential under adding and subtracting an
infinitesimal
fraction, $\omega$, of the integer particle number, $N$:
\begin{equation}
\Delta_{xc} = \lim_{\omega \to 0}\left\{\frac{\delta E_{xc}[n]}{\delta
n}_{|_{N+\omega}} - \frac{\delta E_{xc}[n]}{\delta
n}_{|_{N-\omega}}\right\}_{n_N}.
\end{equation}

Applying LDA to calculate E$_g$ from Eq.~(1), one faces the basic
problem that LDA does not bind the $N+1$ atomic electron system. On the
other hand, using Eq.~(2) within LDA results in the equality of the LDA KS gap and the fundamental gap
since the discontinuity $\Delta_{xc}$ vanishes in this approximation.
As is well known, the LDA-KS absolute gap is always much smaller than
the experimental gap, in some cases, it is even qualitatively wrong.
Namely,  LDA predicts zero gap, a metal, where in nature the system
is a semiconductor.\\[1ex]

A most interesting development in density-functional theory
during last few years was the application of the exact-exchange KS method
(EXX) to the case of crystalline solids.  For this method, the total-energy
functional is given by:
\begin{equation}
E[n] = T_S[n]  +  \int n(r) V_{ext}(r) + \frac{1}{2}\int\int
\frac{n(r) n(r')}{|r-r'|} +E_X[n] + E_C[n],
\end{equation}
where
\begin{equation}
E_X[n] = -
\frac{1}{2}\sum_{kk'\sigma}^{occ}
\int\int\frac{\phi^{*}_{k\sigma}(r)\phi_{k\sigma}(r')\phi_{k',\sigma}(r)
\phi^{*}_{k'\sigma}(r')}{|r-r'|}
\end{equation}
is the exact exchange energy functional.
$T_S[n]$ is the kinetic-energy term of non-interacting electrons and
the correlation-energy term $E_C[n]$ remains to be approximated, 
for example, by LDA.
The orbitals $\phi_k$ in the expression for $E_X$
are KS orbitals, that is the orbitals which
minimize $T_S[n]$ for a given density, $n(r)$, and are therefore functionals
of
the electron
density. For this reason, $T_S$ and $E_X$ are both implicit functionals
of
the density, $n(r)$, and the KS equations can be derived:
\begin{equation}
\left(-\frac{1}{2m}\nabla^2 + V_{ext}(r) + V_H[n](r) + V_X[n](r) + V_C[n](r)\right)
\phi_{k\sigma}(r) = E_{k\sigma} \phi_{k\sigma}(r) ,
\end{equation}
where the KS exchange potential $V_X[n](r)=\delta E_X[n]/\delta
n(r)$
is obtained by the chain-rule differentiation:
\begin{equation}
V_X(r) = \frac{\delta E_X[n]}{\delta n(r)} = \sum_k^{occ}\int\int\left[
\frac{\delta E_X[n]}{\delta \phi_k(r')}\frac{\delta \phi_k(r')}{\delta
V_{KS}(r'')} +
c.c.\right]\frac{\delta V_{KS}(r'')}{\delta n(r)}.
\end{equation}\\
Here $V_{KS}[n](r)$ is the total KS potential corresponding to the
density, $n({\bf r})$.
Neglecting correlation, the
total-energy functional has apparently the form of a Hartree-Fock
expression
for the total energy. However, the EXX method and the
Hartree-Fock method differ because the Hartree-Fock and KS
orbitals $\phi_k$ are not the same.
In the Hartree-Fock method these orbitals obey an equation with a
non-local
potential (the Fock operator), whereas in the KS method they are
determined
by a KS Hamiltonian with a local, multiplicative potential.\\

For more than two decades, the EXX method, sometimes under the name 
\emph{optimized effective potential} (OEP), has been applied to 
atoms and molecules.\cite{TS76,GKKG98} Total ground-state energies were
always found to be extremely close to the Hartree-Fock values while
the single-particle spectrum - yielding, for example, a Rydberg series for
finite systems - is strongly improved over Hartree-Fock.
Only recently, however, has this method
been applied to bulk semiconductors, insulators and 
metals.\cite{K94,K95,KA96,Go,SMV,ASG0,F1} A surprising
result of EXX calculations for sp-semiconductors is that
the KS energy gaps between occupied and unoccupied states are very close to 
experimental gaps. This contrasts typical LDA results in which the
gaps are always too small.  In the EXX case, a larger gap than in LDA is to be 
expected because the EXX potential is
self-interaction free and, thus, binds more strongly than the LDA potential.
For example,
the EXX-KS electron binding energies in atoms are greater in
magnitude than
their LDA counterparts and closer to experiment. Since the occupied
valence
states generally have a greater self-interaction contribution than the
more delocalized unoccupied states, the EXX-KS gap should widen as compared
to the LDA. If the fundamental gap is evaluated from Eq.~(1) with total
energies in Hartree-Fock approximation, one obtains
\begin{equation}
{\rm E}_g^{HF} = \epsilon^{HF}_{N+1}(N) - \epsilon^{HF}_N(N).
\end{equation}
Evaluating, on the other hand, the gap in EXX, Eqs.~(2) and (3) yield
\begin{equation}
{\rm E}_g^{EXX} = \epsilon^{EXX-KS}_{N+1}(N) - \epsilon^{EXX-KS}_N(N)
+ \Delta_x
\end{equation}
where $\Delta_x$ is the X-only discontinuity. Assuming that, like for finite
systems, the total energies in Hartree-Fock and the total energies in EXX 
are very close, one would expect that Eqs.~(9) and (10) yield very similar
values for the band gap. This is indeed the case.\cite{SMV}
What is surprising and still not well understood is why
the EXX-KS gaps alone, i.e. without adding the discontinuity,
are so close to the experimental gaps for sp-semiconductors.
The prevailing belief is that  the \emph{exact} KS gaps are
smaller than the true gaps for solids.  As there are no exact KS
potentials
for solids available, this belief is yet unverified.
It is not known how large the discontinuity, $\Delta_{xc}$, given by Eq.~(4),
of the \emph{exact} exchange-correlation density-functional is.
In any case, the EXX method is the
only KS method so far yielding the electronic structure
close to experiment, provided one does not add the discontinuity which should, 
in principle, be considered. This success has some important practical
consequences.
One of them is the possibility of using the one-electron energy spectrum
as a first, approximate description of excitations and as a fast tool
to interpret experiments. Another, perhaps more important, is
a better starting point for more accurate calculations of excitations via
the time-dependent density-functional theory.\\

In this paper, we explore the performance of the EXX method for materials
very different from sp-semiconductors, namely, the crystalline noble gases.
Solid Ne, Ar, Kr and Xe are special systems since these materials
are composed of almost independent atoms. 
In fact, the shape of the charge density of superimposed
isolated atoms is rather similar to the charge density of the solid
and we might expect the energetics to be similar as well.
The individual tightly bound close-shell
atoms bind very weakly with one another. 
Therefore these systems are a difficult case for LDA.
Indeed, as will be seen in Table IV, the LDA description of the equilibrium 
structural properties for these solids is much less satisfactory than usual.  
These solids are large-gap insulators with their experimental energy gaps 
ranging between 21.4 eV for neon and 9.8 eV for xenon.  Furthermore, they have 
a large exciton binding energy on the order of several eV.
Since the noble-gas solids 
are a loosely bound ensemble of atoms, a comparison between
various electronic properties in the solid phase and in isolated atoms 
is meaningful and interesting.  We will investigate both cases
within the LDA and EXX methods. As will be seen later, in the atomic case the
EXX without correlation resembles highly accurate Kohn-Sham results when 
they are available. Adding LDA correlation to the exact exchange potential 
usually worsens the quality of results. However, for the physical quantities 
involving total-energies differences an account of correlation seems to be 
necessary.\\

Our results show that the EXX approximation, 
with and without LDA correlation,
widens the LDA Kohn-Sham gaps of noble-gas solids by about 1.3 - 3.4 eV.
However, the KS energy gaps are not as close
to the experimental gaps as reported for sp-semiconductors.
Along with previous observations that the EXX-KS gaps for
diamond\cite{SMV} and MgO\cite{K94} are smaller than experimental
gaps, our results suggest that, at least for large-gap insulators,
the EXX theory does not perform as well as it does for
sp-semiconductors.\\

It has been suggested that for atoms the exact KS gap between the
highest occupied and the lowest unoccupied state represents an approximation
of the \emph{optical} gap rather than quasiparticle gap\cite{SUG98,AlRU98}.
The latter represents a nonneutral excitation or a total-energy 
difference between the $N$ and $N\pm1$ particle systems while the former
is the gap between the ground and excited states of the $N$-electron system.
This conjecture has been explicitly verified in the cases where
\emph{almost exact} Kohn-Sham atomic potentials have been 
calculated\cite{SUG98,AlRU98}.
Our EXX results for noble-gas atoms (Section II) support this conjecture 
as well.  In regard to the solids, the picture is not as clear.
For semiconductors,
the exciton binding energy is very small, of the order of meV, the
observed agreement between the EXX-KS gaps and experimental gaps
could support the conjecture of Refs.~[\onlinecite{SUG98,AlRU98}] as well.
However, unless better approximations to the correlation-energy functional
are known this remains as a rather inconclusive speculation.
Our results for noble-gas solids, systems where optical and
quasiparticle gaps differ by several eV and where electronic structure
in the solid phase is similar to the structure of energy levels in isolated 
atoms, could shed some light on the validity of this conjecture.
What we have found is that the EXX gaps for noble-gas solids are -
unlike the case of semiconductors - significantly smaller than the 
quasi-particle gaps. However, they are also smaller than the 
experimental optical gaps, reproducing about 80 \% of their values.
A fundamental question remains then: would an \emph{exact} treatment of correlation
align the KS and optical gaps for noble-gas solids as well, or does
correlation in the solid phase have a qualitatively different character
and role than for finite, atomic systems, so that the conjecture
of Refs.~[\onlinecite{SUG98,AlRU98}] is not correct for solids ?\\

The next section is devoted to results for isolated atoms. Electronic 
properties of the noble-gas solids are presented and discussed in Section III.

\section{Noble-gas atoms: Ne, Ar, Kr and Xe}

Since noble-gas solids consist of loosely bound individual atoms,
it is instructive to start the investigation
by analyzing the properties of isolated atoms. In addition,
the density-functional methods for atoms are more
advanced than in extended systems and can be used as a test for
various approximations. This is for two reasons. First,
highly accurate KS potentials are available for a few
light atoms like Ne\cite{UG93,ZMP94} and Ar\cite{ZMP94}. Second, for finite
systems it is
computationally feasible to calculate the fundamental gap, Eq.~(1), from
total-energy differences. Note that for finite systems, the fundamental gap
as given by Eq.~(1) is usually called the chemical hardness. For simplicity,
we keep the term ``fundamental gap'' for both finite and infinite systems.
In the following we present several results
for the noble-gas atoms and draw some conclusions about the validity
of certain approximations.\\

In Tables~I and II we present the KS eigenvalues for the Ne and the Ar atom,
respectively. Results using LDA, pure EXX (EXX),
EXX plus LDA correlation (EXXc) and highly accurate (\emph{almost exact})
KS potentials for Ne and Ar atoms are presented.
The column denoted QMC in Table~I shows eigenvalues resulting from the KS
potential obtained by Umrigar and Gonze\cite{UG93} through a Quantum Monte
Carlo calculation of the ground-state density of the Ne atom. The column 
denoted CI in Tables I and II shows the results obtained by Morrison and 
Zhao\cite{MZ95}
from highly accurate densities which were calculated through the
configuration-interaction approach.  The column called \emph{Expt.} contains
the negative of the experimental first
ionization potential of each
atom which should be compared to the uppermost occupied KS eigenvalue.
The column also contains the experimental binding energies of a few excited
states. The latter are defined as the negative of the difference
between the experimental ionization potential and the experimental optical
energy gap between the ground state and the relevant excited state.
For the optical gap we take a weighted average of singlet and triplet gaps.
As for the occupied levels, we report only the experimental ionization 
potential, i.e. the binding energy of the highest occupied state.
This is because it is unclear how the lower KS eigenvalues compare with the
physical binding energies.
However, for unoccupied states, it has been
suggested that the KS eigenvalues are a good approximation to 
the experimental binding energies of the excited atoms \cite{SUG98,AlRU98}.
This idea has been explicitly verified for helium\cite{SUG98},
beryllium\cite{SUG98} and neon atoms\cite{AlRU98}; and some plausible
arguments have been put forward to explain why.\cite{SUG98}\\

Tables~I and II show that the LDA KS
eigenvalues are smaller in magnitude
than their EXX and EXXc counterparts. The magnitude is largely a 
consequence of the self-interaction error inherent in the LDA method.
Another consequence of this self-interaction error in LDA  
is the absence of none higher but the first bound, 
unoccupied state. These facts are well known. What is more interesting 
for this study is that EXX eigenvalues for Ne and Ar are deeper than the exact ones.
These facts are well known. What is more interesting
for us is that EXX eigenvalues for Ne and Ar are deeper than the exact ones.
This means that the EXX method binds atomic states of Ne and Ar too
strongly and exact correlation would correct for this over-binding. When LDA
correlation are added to the exact exchange (EXXc method) the situation 
worsens. 
For example, the EXX uppermost occupied eigenvalues are roughly
1.5, 0.3 eV deeper than the exact eigenvalues for Ne and Ar respectively. 
In the EXXc method the difference grows to 3.2 and 1.7 eV respectively. 
It is interesting to note that for occupied states this expected behavior 
of \emph{exact} correlation of the KS theory is similar to the role of 
correlation in the many-body Green's functions approach. There, the many-body 
correlation shift the eigenvalues, or rather the poles of the 
one-particle Green's functions, in the opposite direction as the Hartree-Fock 
shift. The well known effect is that
binding energies of occupied states are much too strong within
Hartree-Fock. It turns out that the EXX KS occupied states are also more
bound than the expected exact KS states, but the effect is much
smaller than in the Hartree-Fock theory.
The analogy brakes down for empty states. In this case,
EXX results in still deeper unoccupied states than
in experiment and adding LDA correlation makes them even deeper.  On the other hand, Hartree-Fock theory hardly binds at all. This striking difference between
the EXX and Hartree-Fock theories is because Hartree-Fock only has an asymptotic $-1/r$ potential for
the occupied states and has an exponentially decaying potential
for the empty states. In contrast,
the KS $V_{XC}$ potential has a Coulombic tail for all states.
One says, that the Hartree-Fock potential is self-interaction free only
for occupied states, whereas the KS potential (exact and EXX)
is self-interaction free for all states.\\

For the heavier atoms Kr and Xe, we perform a full
relativistic OEP calculation as relativistic effects should be important.
In these cases, we do not have exact KS potentials or
eigenvalues with which to compare the results. However, a comparison with the
experimental first ionization potential shows that, when the spin-orbit
interaction is neglected, the exact exchange
calculation gives a slightly over-bound uppermost occupied level.
Here again, LDA correlation lower the eigenvalues
too far. It is clear that for all noble-gas atoms adding
LDA correlation to exact KS exchange deteriorates the one-electron
properties.\\

So far we have compared LDA, EXX and EXXc KS eigenvalues
with the \emph{exact} ones, where available, and with the experimental
first
ionization potential and binding energies of unoccupied states.
For finite systems, it is possible to relate certain excitations
with total-energy differences. This is the case for the first ionization
potential $I=E[N-1]-E[N]$, electron affinity $A=E[N]-E[N+1]$, and the
energy gap E$_g$, Eq.~(1).
The electron affinity, $A$, as defined by a total energy difference
is zero within LDA theory for most atoms.  This
is because in LDA the corresponding $N+1$ electron system is not bound.
In the case
of noble-gas atoms, however, the experimental affinity does vanish, and
the LDA result is fortuitously correct.  Since the affinity vanishes, the gap 
must equal the ionization energy. The same argument is valid
for EXX.  On the other hand, EXXc gives a small but
finite value for the affinity.
Table III shows the atomic energy gaps E$_g$, both experimental and
calculated
from total-energy differences within LDA, EXX and EXXc.
These energy gaps are compared to the KS eigenvalue differences
$\epsilon_{N+1}(N) - \epsilon_N(N)$.
In addition, the (triplet-singlet
averaged) first optical gap is presented $\bar{\Delta}_{opt}$,
as well as a calculated in each
method \emph{total-energy optical gap} $\Delta^*=E[N]^*-E[N]$. 
Here. $E[N]$
is the total ground-state energy and $E[N]^*$ is the self-consistent 
total energy of an excited state
in which there is a hole in the uppermost p-shell and an electron in the
next s-shell. \\

First, we note that the KS gap rather poorly approximates the
experimental fundamental gap.  The average deviation over the four elements
between the KS and true gaps is 5.9, 4.0, 3.1 eV for LDA, EXX and
EXXc respectively.  As shown by the data, LDA KS gaps deviate
the most. When expressed in percents, LDA, EXX and EXXc KS gaps
account for 63 \%, 75 \% and 80 \% of the experimental fundamental atomic gaps
respectively. For Ne~(Table~I), the highly accurate 
KS energy gap amounts to 77 \% of the experimental one. For Ar~(Table~II)
we could only estimate the exact KS gap to be of about 68 \%
of the experimental one. The situation changes when we compare KS
gaps to experimental optical gaps.
It turns out that LDA, EXX and EXXc reproduce
the atomic optical gaps with 13 \%, 3 \% and 9 \% accuracy respectively.
The highly accurate KS gap of Ne amounts to 99.9 \% of the optical gap.
In particular, the good agreement of the EXX method with experiment is to be 
noted. An interesting question is whether similar trends take place in the 
solid phase.  We might expect the answer to be yes if the solid is composed of 
a bunch of weakly interacting atoms.\\

When we compare the experimental fundamental gap of noble-gas atoms with the
calculated total-energy differences (Eq.~1) the agreement is fairly good.
Across four elements, LDA, EXX and EXXc
reproduce the experimental atomic
gaps on average to
4 \%, 6 \% and 3 \% accuracy respectively.
Performing a similar
although less rigorously justified total-energy calculation of
\emph{optical gaps} the agreement amounts to 5 \%, 5 \% and 4 \% for LDA,
EXX and EXXc respectively.
What is remarkable is
that for a total-energy difference calculations
accounting for correlation seems
to be very important.  In fact, the best total-energy difference
results come from the EXXc method, the worse with the pure EXX one.\\

In the next section we will investigate whether similar trends take place
in the solid phase.\\[2ex]

\section{Noble-gas solids: Ne, Ar, Kr and Xe}

We performed our calculations for noble-gas solids within the
pseudopotential and
plane-wave formalism. For each approximate $E_{xc}$ (LDA, EXX and EXXc),
a pseudopotential was generated using the same functional \cite{Mou}. 
We followed the EXX plane-wave formalism
developed by G\"orling\cite{Go}
and St\"adele {\emph et al.}\cite{SMV}.  This formalism had been 
applied in our previous publication \cite{F1}.
The plane-wave cutoff ranged between 50 and 120 Rydbergs
depending on the material and whether the structural or electronic
properties were investigated. For solid krypton and xenon,
the spin-orbit interaction was taken into account perturbatively.\\

In Table~IV the experimental and calculated equilibrium fcc lattice
constants are presented. The percent deviations
from experiment are also given. As already noted, noble-gas solids
resemble loosely bound isolated atoms. For such systems the total energy
only weakly depends on the interatomic distance.  The energy-volume
curve is very flat and the system hardly binds. In the
absence of stronger interatomic interactions, a theoretical determination
of equilibrium properties is subject to a rather large uncertainty.
Usual convergence criteria for total-energy determination can easily become
insufficient and minor computational details, like for example, 
details of the
pseudo-potential construction can matter. For all these reasons we have
checked our LDA results against independent results from publicly 
available LDA codes.\cite{ABINIT,fhi98md}\\

As shown
in Table~IV, the structural equilibrium properties of noble-gas solids
are very poorly described by the all methods. For example,
the error in the estimate of the lattice constant for neon
is more than 13 \%. This discrepancy is unusually large; 
for most solids, LDA gives a lattice-constant estimate
within about 1-2 \%. For Ar, Kr and Xe, the LDA discrepancy is on 
the order of 5 \%. Although not as dramatic as for Ne, this
disagreement is still much larger than usual.
Using the EXX method
without correlation does not help much; the disagreement with experiment is
just as large as in the other two cases. 
In contrast to the LDA which underestimates
the lattice spacing, the EXX method overestimates it for Ar, Kr and Xe.
With the exception of neon, the combined method, exact-exchange plus LDA 
correlation (EXXc), is the closest to experiment, but it is still not perfect.
This result is similar to what was seen in the previous section's 
atomic calculations. For total energy differences,
the EXXc method gives the most accurate results. It is worth noting,
that none of these approximations properly account for the long-range 
Van der Waals interactions that are responsible
for the binding of noble-gas solids. It is not surprising that 
structural properties differ so much from experiment.\\

Figures I-IV show the band structures along the L-$\Gamma$-X directions
for Ne, Ar, Kr and Xe with Kr and Xe being treated relativistically. 
The solid lines represent the EXX band structure,
the dashed lines correspond to the LDA one. In Tables~V-VIII the KS
energies
at high symmetry points $\Gamma$, X and L are presented. In Table~IX
the KS energy gaps are compared with the experimental absolute
energy gaps and with the corresponding optical gaps for solid Ne, Ar, Kr and Xe.
It is important to keep in mind that for this special class of elements, the
description of the electronic structure provided by LDA, EXX and EXXc
might be expected to work less accurately in the solid phase
than for isolated atoms.
This is because in addition to only approximating short-range correlation,
our functionals do not account for the long-range Van der Waals effects.\\

Nevertheless, it is interesting to consider how well LDA, EXX and EXXc
describe the electronic structure of noble-gas solids.
In LDA, the KS gaps are on average 55 \% of the experimental gaps. 
This is a typical result for solids, and
though slightly less accurate than for the atomic limit given by LDA.
For noble-gas solids EXX and EXXc reproduce about 68 \% of the fundamental gap,
whereas in the atomic limit the KS gap ranged on average to 74 \% and 79 \%
of the experimental fundamental gap for EXX and EXXc respectively. 
The fact that the EXX and EXXc KS gaps in noble-gas solids are not very close
to the experimental fundamental gaps is a central result of our
investigation. The atomic results suggested that the EXX-KS gaps should be 
close to the optical gaps. Since the noble-gas solids have
a large exciton binding energy, the optical gap is appreciably smaller
than the fundamental gap. What we observe in Table~IX is that the EXX
and EXXc KS gaps are still significantly smaller than optical gaps in 
noble-gas solids. They amount to about 80 \% - 81 \% of the experimental
al-gaps, whereas the agreement is 97 \% in the atomic limit.\\

The valence bands show very little dispersion in Figs.~1 through 4. 
Furthermore, the energy distance between occupied s and p 
bands is very close to the KS energy gap between s and p valence states
in the isolated atoms. This confirms the popular picture that rare-gas
solids are composed of almost undisturbed atoms. However, the conduction
bands exhibit a rather pronounced dispersion. It would be interesting
to verify this result experimentally. Clearly, an indirect confirmation
is the fact that the experimental optical gap of the solid is always
larger than the optical gap in the corresponding atom: The hole and the
electron attract each other strongly if they are both localized on the
same atom. In the solid, on the other hand, the hole is similarly
localized as in the atom, while the electron in the conduction band
is more delocalized, leading to a weaker interaction with the hole.\\

Another remarkable result is that the total valence-band width resulting
from LDA is smaller than the one from EXX for
neon. For argon, both widths are comparable.  For krypton
and xenon, the EXX width is smaller. We recall that previous EXX calculations
have shown that for sp-semiconductors, EXX leads to narrower
total valence-band widths than in LDA. 
Only for diamond was the opposite observed.\\

\section{Conclusions}

We have applied the exact exchange method within density-functional theory
to the noble-gas solids, Ne, Ar, Kr and Xe. It was previously shown for 
He, Be and Ne atoms that the Kohn-Sham energy gap coming from a nearly exact
KS potential is an excellent approximation to the atomic optical gap but not 
to the fundamental (quasiparticle) gap. The EXX-KS gaps for these atoms are
also in very good agreement (3 \% in average) with experimental optical gaps.
A central question of our investigation was whether the same holds for
the noble-gas solids. It turns out that, in contrast to previous results
for sp-semiconductors, the EXX-KS gaps in noble-gas solids are
appreciably smaller than the experimental fundamental gaps. Moreover,
they are also smaller by 20\% than the experimental optical gaps.
The results of our investigation clearly show that the EXX method
does not provide a KS band structure that agrees equally well
with experiment for semiconductors and insulators.\\[2ex]

\section{Acknowledgments}

We would like to thank J.~A.~Majewski and M.~Moukara for making available
their EXX-pseudopotential-generation code \cite{Mou} and  U.~von Barth
for valuable discussions. We thank X.~Gonze for making available the 
{\em nearly exact} KS potential of the Ne atom.
Financial support from the Deutsche 
Forschungsgemeinschaft (SFB 410) is acknowledged (A.~Fleszar).
R.~Magyar was funded by a grant from the German Academic Exchange Service
(DAAD) and under NSF grant CHE-9875091. This work was supported in part
by the EXC!TING Network of the EU.
Some calculations were performed at the Forschungszentrum J\"ulich,
part at the HLRS Stuttgart.

\begin{table}[p]
\begin{tabular}[t]{crrrrrr}
Ne & \multicolumn{1}{c}{LDA} & \multicolumn{1}{c}{EXX} &
\multicolumn{1}{c}{EXXc} & \multicolumn{1}{c}{QMC} &
\multicolumn{1}{c}{CI} & \multicolumn{1}{c}{Expt.}\\[2ex]
\hline
1s & -824.34 & -838.30 & -840.38 & -838.18 & -838.30 & \\
2s &  -35.97 &  -46.73 &  -48.40 &  -44.93 &  -45.01 & \\
2p &  -13.54 &  -23.14 &  -24.76 &  -21.61 &  -21.69 & -21.56 \\
\hline
3s &   -0.07 &   -5.23 &   -5.77 &   -4.97 & & -4.9 \\
3p &         &   -3.11 &   -3.40 &   -3.00 & & -2.94 \\
4s &         &   -1.95 &   -2.03 &   -1.90 & & -1.89 \\
3d &         &   -1.57 &   -1.63 &   -1.55 & & -1.53 \\
\end{tabular}
\vspace{1em}
\caption{Neon-atom energy levels (in eV).  The Kohn-Sham energies are from LDA,
exact exchange (EXX) and exact exchange with LDA correlation
(EXXc). Column QMC gives the eigenvalues obtained with the almost
exact Kohn-Sham potential of Ref.~[18]. Column CI presents
results of almost exact Kohn-Sham calculation of Ref.~[20].}
\label{tab1}
\end{table}

\begin{table}[p]
\begin{tabular}[t]{crrrrr}
Ar & \multicolumn{1}{c}{LDA} & \multicolumn{1}{c}{EXX} &
\multicolumn{1}{c}{EXXc} & \multicolumn{1}{c}{CI} &
\multicolumn{1}{c}{Expt.} \\[2ex]
\hline
1s & -3095.39 & -3112.99 & -3115.42 & -3113.82 & \\
2s &  -293.61 &  -303.27 &  -305.13 &  -302.59 & \\
2p &  -229.67 &  -237.46 &  -239.36 &  -236.85 & \\
3s &   -24.02 &   -29.90 &   -31.37 &   -28.79 & \\
3p &   -10.40 &   -16.07 &   -17.48 &   -14.88 & -15.76 \\
\hline
4s &    -0.26 &    -4.37 &    -4.94 & & -4.08  \\
4p &          &    -2.77 &    -3.09 & & -2.66  \\
3d &          &    -1.86 &    -2.29 & & -1.83  \\
\end{tabular}
\vspace{1em}
\caption{Argon-atom energy levels (in eV).  The Kohn-Sham energies are from LDA,
exact exchange (EXX) and exact exchange with LDA correlation
Column CI presents results of almost exact Kohn-Sham calculation of 
Ref.~[20].}
\label{tab2}
\end{table}

\begin{table}[p]
\begin{tabular}[t]{llrrrr}
 & & \multicolumn{1}{c}{Ne} & \multicolumn{1}{c}{Ar} &
\multicolumn{1}{c}{Kr} & \multicolumn{1}{c}{Xe} \\[2ex]
\hline
Expt: & E$_g$                & 21.56 & 15.76 & 14.00 & 12.13 \\
Expt: & $\bar{\Delta}_{opt}$ & 16.63 & 11.57 &  9.94 &  8.35 \\
\hline

LDA:  & E$_g$           & 22.66 & 16.17 & 14.44 & 12.73 \\
      & $\Delta^*$      & 17.74 & 11.96 & 10.32 &  8.87 \\
 &$\Delta\epsilon_{Kohn-Sham}$  & 13.47 & 10.14 &  8.76 &  7.50 \\[1ex]
\hline

EXX:  & E$_g$           & 19.83 & 14.77 & 13.22 & 11.66 \\
      & $\Delta^*$      & 15.16 & 10.96 &  9.58 &  8.32 \\
 &$\Delta\epsilon_{Kohn-Sham}$  & 17.91 & 11.70 &  9.81 &  8.12 \\[1ex]
\hline

EXXc: & E$_g$           & 21.31 & 16.03 & 14.61 & 12.79 \\
      & $\Delta^*$      & 16.08 & 11.72 & 10.28 &  8.97 \\
 &$\Delta\epsilon_{Kohn-Sham}$ & 18.99 & 12.54 & 10.58 &  8.82 \\

\end{tabular}
\vspace{1em}
\caption{Fundamental energy gaps E$_g$=I-A and optical gaps from experiment
and calculations in neutral atoms Ne, Ar, Kr and Xe. $\bar{\Delta}_{opt}$ is 
the multiplet-averaged experimental transition energy from the ground state 
to $p^5s^1$ state. $\Delta^*$ is the calculated total-energy difference between
the excited atom in the ($p^5s^1$) configuration and the ground state. 
$\Delta\epsilon_{Kohn-Sham}$ is the Kohn-Sham gap.}
\label{tab3}
\end{table}

\begin{table}[p]
\begin{tabular}[t]{lrrrr}
 & Ne & Ar & Kr & Xe \\[2ex]
\hline
$a^{Expt}$   & 8.44    & 9.94   & 10.66  & 11.59 \\[1ex]

$a^{LDA}$ & 7.29      & 9.35    & 10.13   & 11.14    \\[-1ex]
          & 13.6 \% & 5.9 \%& 5.0 \%& 3.9 \% \\[1ex]
$a^{EXX}$ & 7.23      & 10.13   & 11.07   & 12.66   \\[-1ex]
          & 14.3 \% & 1.9 \%& 3.8 \%& 9.2 \%\\[1ex]
$a^{EXXc}$& 7.06      & 9.80    & 10.77   & 12.06    \\[-1ex]
          & 16.4 \% & 1.4 \%& 1.0 \%& 4.1 \% \\[1ex]

\end{tabular}
\vspace{1em}
\caption{ Equilibrium cubic lattice spacing (in a.u.) 
from experiment and calculations. Percents 
show the diviations of the lattice constant from experiment.}
\label{tab4}
\end{table}

\begin{table}[p]
\begin{tabular}[t]{crrr}
Ne & \multicolumn{1}{c}{LDA} & \multicolumn{1}{c}{EXX} &
   \multicolumn{1}{c}{EXXc} \\[2ex]
\hline
$\Gamma$  & -22.98 & -24.04 & -24.22 \\
          &   0.00 &   0.00 &   0.00 \\
          &  11.32 &  14.15 &  14.76 \\
          &  29.26 &  31.71 &  32.28 \\[1.5ex]

X         & -22.87 & -23.96 & -24.14  \\
          &  -0.62 &  -0.56 &  -0.53 \\
          &  -0.20 &  -0.22 &  -0.21 \\
          &  18.21 &  21.24 &  21.91 \\
          &  19.00 &  21.79 &  22.38 \\[1.5ex]

L         & -22.90 & -23.98 & -24.16  \\
          &  -0.69 &  -0.58 &  -0.55 \\
          &  -0.07 &  -0.06 &  -0.06 \\
          &  17.06 &  19.84 &  20.42 \\
          &  17.21 &  20.05 &  20.68 \\
\end{tabular}
\vspace{1em}
\caption{Ne-solid Kohn-Sham eigenvalues in eV from LDA, EXX and EXX plus LDA
correlation (EXXc) at high symmetry points.}
\label{tab5}
\end{table}

\begin{table}[p]
\begin{tabular}[t]{crrr}
Ar & \multicolumn{1}{c}{LDA} & \multicolumn{1}{c}{EXX} &
   \multicolumn{1}{c}{EXXc} \\[2ex]
\hline
$\Gamma$  & -14.57 & -14.48 & -14.51 \\
          &   0.00 &   0.00 &   0.00 \\
          &   8.16 &   9.61 &  10.14 \\
          &  15.51 &  16.01 &  16.37 \\
          &  17.89 &  18.08 &  18.37 \\[1.5ex]

X         & -14.28 & -14.20 & -14.25  \\
          &  -1.27 &  -1.14 &  -1.06 \\
          &  -0.45 &  -0.42 &  -0.39 \\
          &  10.85 &  12.02 &  12.57 \\
          &  12.34 &  13.24 &  13.70 \\
          &  14.89 &  16.31 &  16.67 \\[1.5ex]

L         & -14.35 & -14.27 & -14.32  \\
          &  -1.40 &  -1.25 &  -1.16 \\
          &  -0.15 &  -0.14 &  -0.14 \\
          &  11.03 &  12.17 &  12.65 \\
          &  13.29 &  14.80 &  15.24 \\
          &  15.12 &  15.69 &  16.09 \\
\end{tabular}
\vspace{1em}
\caption{Ar-solid Kohn-Sham eigenvalues in eV from LDA, EXX and EXX plus LDA
correlation (EXXc) at high symmetry points.}
\label{tab6}
\end{table}

\begin{table}[p]
\begin{tabular}[t]{crrr}
Kr & \multicolumn{1}{c}{LDA} & \multicolumn{1}{c}{EXX} &
   \multicolumn{1}{c}{EXXc} \\[2ex]
\hline
$\Gamma$  & -14.81 & -13.74 & -14.62 \\
          &  -0.73 &  -0.65 &  -0.67 \\
          &   0.00 &   0.00 &   0.00 \\
          &   6.47 &   7.87 &   8.02 \\
          &  13.18 &  13.43 &  13.80 \\[1.5ex]

X         & -14.55 & -13.45 & -14.39  \\
          &  -1.77 &  -1.65 &  -1.59 \\
          &  -0.88 &  -0.82 &  -0.79 \\
          &  -0.54 &  -0.52 &  -0.49 \\
          &   8.76 &   9.65 &  10.05 \\
          &  10.09 &  10.70 &  11.11 \\
          &  13.28 &  14.49 &  14.78 \\
          &  16.94 &  17.63 &  17.96 \\[1.5ex]

L         & -14.62 & -13.53 & -14.45  \\
          &  -1.88 &  -1.74 &  -1.66 \\
          &  -0.59 &  -0.54 &  -0.54 \\
          &  -0.18 &  -0.18 &  -0.17 \\
          &   8.92 &   9.88 &  10.16 \\
          &  11.79 &  13.07 &  13.41 \\
          &  12.77 &  13.09 &  13.47 \\
\end{tabular}
\vspace{1em}
\caption{Kr-solid Kohn-Sham eigenvalues in eV from LDA, EXX and EXX plus LDA
correlation (EXXc) at high symmetry points.  
Spin-orbit splittings included.}
\label{tab7}
\end{table}

\begin{table}[p]
\begin{tabular}[t]{crrr}
Xe & \multicolumn{1}{c}{LDA} & \multicolumn{1}{c}{EXX} &
   \multicolumn{1}{c}{EXXc} \\[2ex]
\hline
$\Gamma$  & -13.01 & -11.14 & -12.71 \\
          &  -1.44 &  -1.27 &  -1.34 \\
          &   0.00 &   0.00 &   0.00 \\
          &   5.26 &   6.69 &   6.51 \\
          &  10.07 &   9.96 &  10.50 \\[1.5ex]

X         & -12.70 & -10.71 & -12.41  \\
          &  -2.43 &  -2.29 &  -2.24 \\
          &  -1.20 &  -1.15 &  -1.09 \\
          &  -0.67 &  -0.66 &  -0.61 \\
          &   6.53 &   7.01 &   7.47 \\
          &   7.43 &   7.64 &   8.17 \\
          &  11.66 &  12.62 &  12.84 \\
          &  14.18 &  13.85 &  14.42 \\[1.5ex]

L         & -12.77 & -10.81 & -12.48  \\
          &  -2.48 &  -2.34 &  -2.27 \\
          &  -0.93 &  -0.86 &  -0.85 \\
          &  -0.23 &  -0.23 &  -0.21 \\
          &   6.85 &   7.53 &   7.78 \\
          &   9.73 &   9.68 &  10.22 \\
          &   9.78 &   9.73 &  10.28 \\
\end{tabular}
\vspace{1em}
\caption{Xe-solid Kohn-Sham eigenvalues in eV from LDA, EXX and EXX plus LDA
correlation (EXXc) at high symmetry points.  
Spin-orbit splittings included.}
\label{tab8}
\end{table}

\begin{table}[p]
\begin{tabular}[t]{crrrrr}
eV & \multicolumn{1}{c}{E$_g^{LDA}$} & \multicolumn{1}{c}{E$_g^{EXX}$} &
   \multicolumn{1}{c}{E$_g^{EXXc}$} & \multicolumn{1}{c}{E$_g^{Expt}$} &
  \multicolumn{1}{c}{$\Delta^{Expt}$}  \\[2ex]
\hline
   Ne     & 11.32 & 14.15 & 14.76 & 21.4 & 17.4 \\[1.5ex]
   Ar     &  8.16 &  9.61 &  9.95 & 14.2 & 12.2 \\[1.5ex]
   Kr     &  6.47 &  7.87 &  8.02 & 11.6 & 10.2 \\[1.5ex]
   Xe     &  5.26 &  6.69 &  6.51 &  9.8 &  8.4 \\[1.5ex]

\end{tabular}
\vspace{1em}
\caption{Calculated and measured energy gaps in noble-gas solids in eV. 
E$_g^{LDA}$, E$_g^{EXX}$
and E$_g^{EXXc}$ are Kohn-Sham gaps from LDA, pure EXX and EXX plus LDA
correlation respectively. E$_g^{Expt}$ is the experimental fundamental
gap.  $\Delta$ is the experimental optical gap.}
\label{tab9}
\end{table}

\newpage
\begin{center} FIGURES
\end{center}
\begin{tabular}{p{2cm}p{11cm}}

Fig.\ 1. & Band structure of Ne along L-$\Gamma$-X directions calculated
within EXX (solid lines) and LDA (dashed lines).\\

Fig.\ 2. & Band structure of Ar along L-$\Gamma$-X directions calculated
within EXX (solid lines) and LDA (dashed lines).\\

Fig.\ 3. & Band structure of Kr along L-$\Gamma$-X directions calculated
within EXX (solid lines) and LDA (dashed lines). Spin-orbit splitting
included.\\

Fig.\ 4. & Band structure of Xe along L-$\Gamma$-X directions calculated
within EXX (solid lines) and LDA (dashed lines). Spin-orbit splitting
included.\\

\end{tabular}

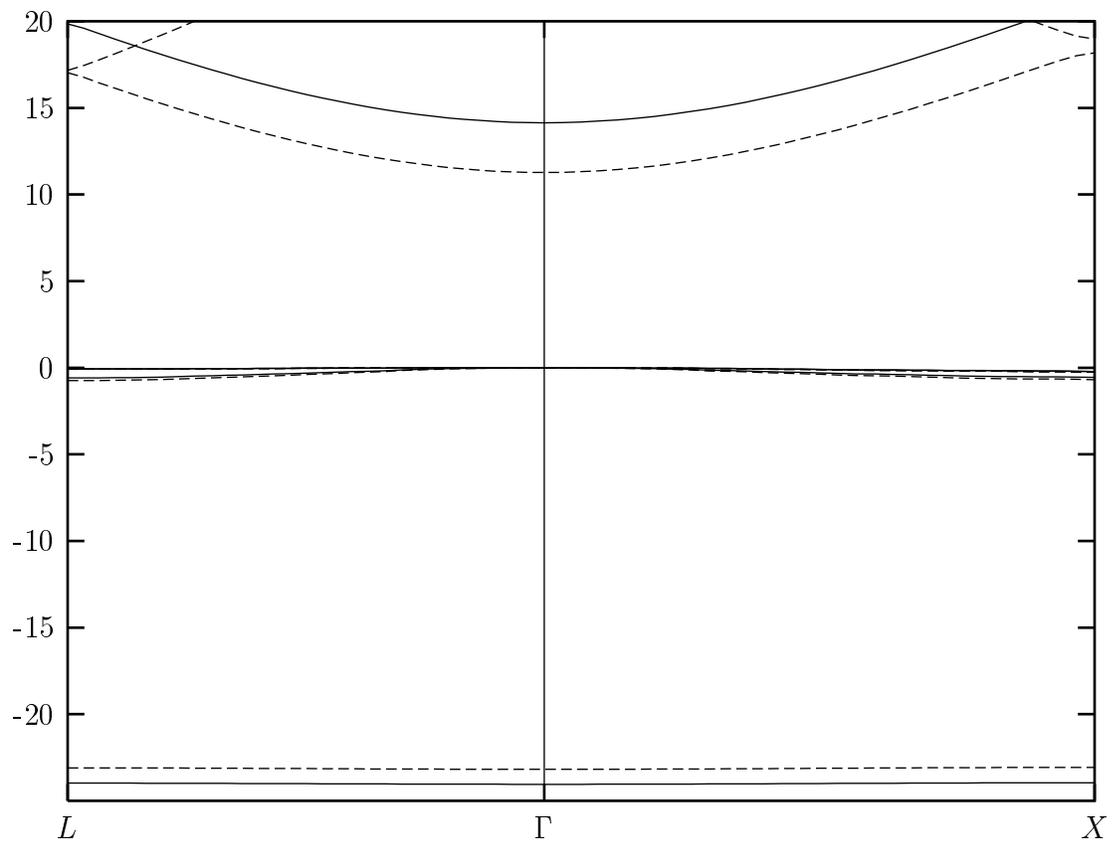
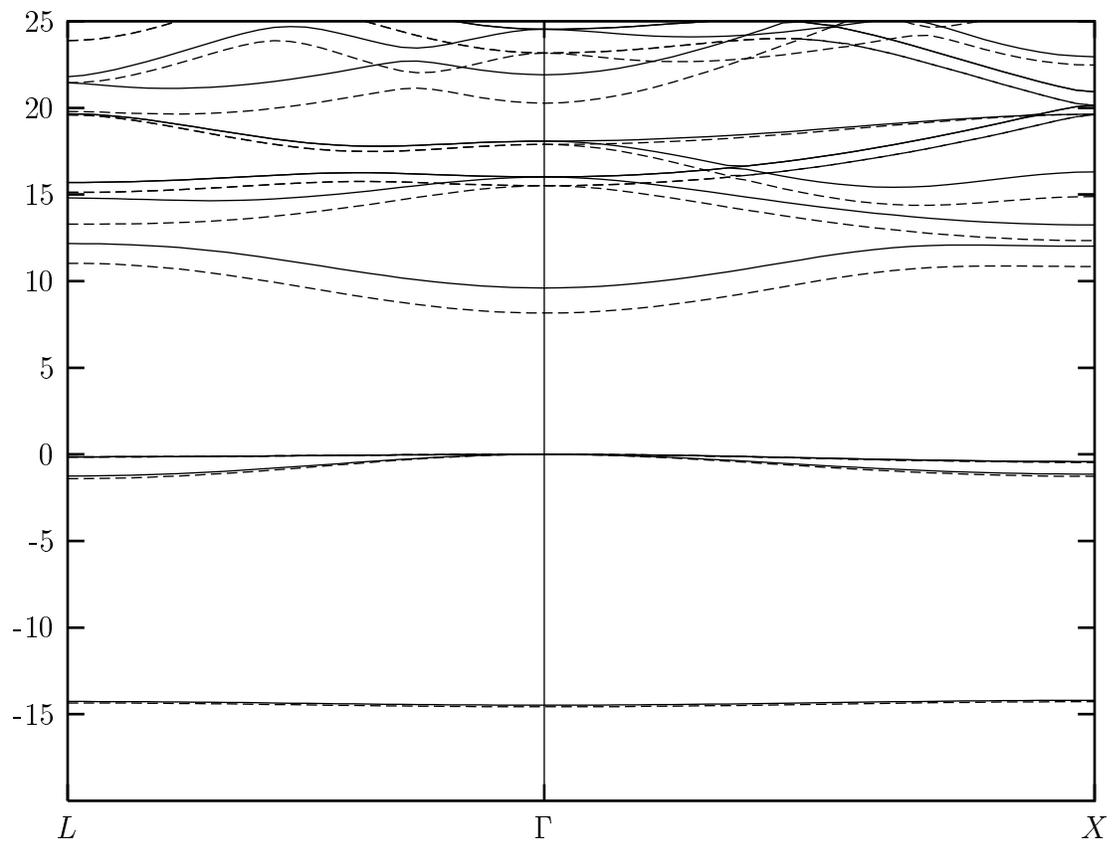
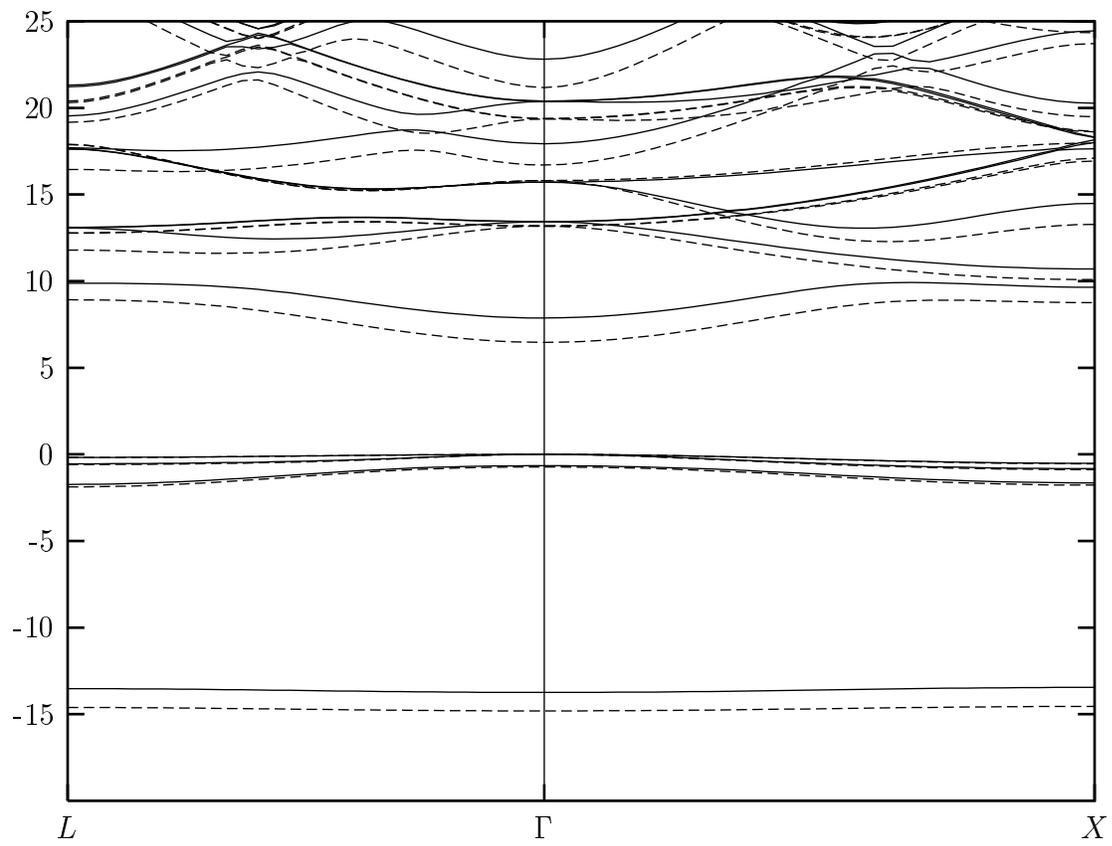
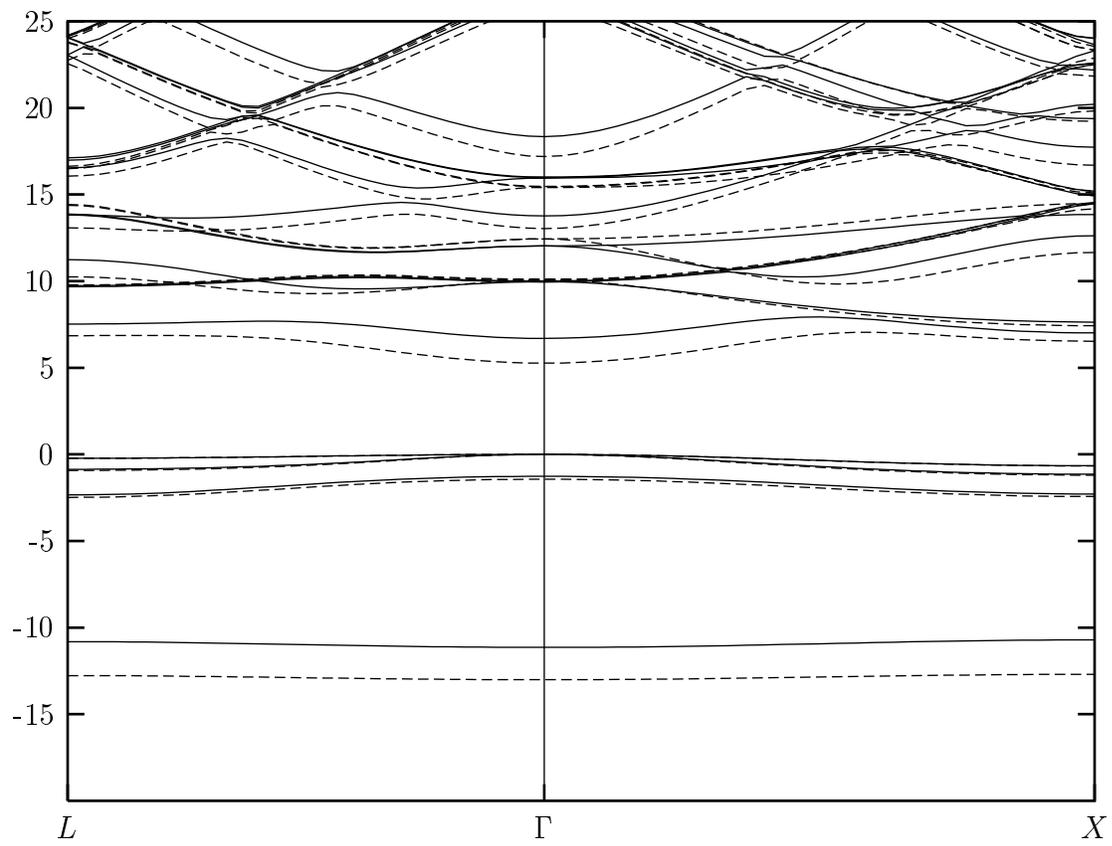
\end{document}